# Enhanced Second-harmonic Generation Using Nonlinear Metamaterials

Sinhara R Silva, Khagendra Battarai, Alexander Shields, Jiangfeng Zhou

University of South Florida

Abstract

We demonstrate a nonlinear metamaterial which enhance higher order harmonics in microwave frequency regime. Nonlinearity in the structure is introduced by adding a varactor diode in the common slit of the double split ring resonator (DSRR) design. By engineering the structure such that inner ring resonance frequency of the DSRR is twice as the outer ring resonance frequency, we have demonstrated that the second harmonic of the outer ring can be enhanced significantly. By comparing with a single ring (SRR) unit cell structure, DSRR has an enhancement factor of 70. Furthermore, we elucidate that the second harmonic signals generated by two identical double split rings interfere and we can use its constructive interference positions to construct an array of DSRRs that gives a maximum second harmonic signal with phase matching condition. When the phase matching condition occurs the enhancement factor of an array is more than the contribution of individual unit cells which proves that the enhancement is due to not only constructive interference of the second harmonic generated by individual unit cells but also due to cavity mechanism occurred in the array structure.

Introduction

A great deal of research in the field of metamaterial was conducted in a linear regime, where the electromagnetic responses are independent of the external electric or magnetic fields. Unfortunately, in linear regime the desired properties of metamaterials have only been achieved within a narrow bandwidth, around a fixed frequency. Therefore, nonlinearity is introduced into metamaterials by merging meta-atoms with well-known nonlinear materials. Among different ways to introduce nonlinearity to the metamaterial, embedding an additional controllable media[1, 2] or inserting nonlinear elements[3-5] to the structure has led to extensive studies on nonlinear metamaterials and demonstrated novel ways to obtain nonlinear responses in metamaterials. Studies have shown nonlinear metamaterials show fascinating features such as resonant frequency tuning[6-8], bistability[9, 10], harmonic generation[11-13], parametric amplification[14] , modulation[15] and backward phase matching[16] .

Among basic metamaterial structures, split ring resonator (SRR) as shown in fig 1(a) is known as the most common, best characterized[9, 17-20] metamaterial with geometrically scalable meta-atoms that can be translated to operability in many regimes of frequencies. Recent studies[21] have shown that nonlinear resonating structures such as SRRs can be advantageous to use in nonlinear metamaterial designs since it has been observed that Split ring resonator produces a huge electric field at its gap at the resonance even it is exposed to an low incident electromagtic waves. In order to increase the nonlinear polarization generated by a non linear medium we can either increase suseptibility $\chi^{(n)}$ or increase the applied electric field in the medium which explain by the nonlinear polarization equation (1).

Wave equation for Propagation of Electromagnetic waves in a nonlinear medium

$$\nabla^2\epsilon - \frac{1}{c^2}\frac{\partial^2\epsilon}{\partial t^2} = \mu_0\frac{\partial^2 P}{\partial t^2}$$

Polarization density

$$P = \varepsilon_0\left(\chi^{(1)}\epsilon + \chi^{(2)}\epsilon^2 + \chi^{(3)}\epsilon^3 + \cdots\right)$$

$$P = \varepsilon_0\chi^{(1)}\epsilon + P_{NL}$$

Induced polarization is given by

$$P_{NL} = \varepsilon_0\left(\chi^{(2)}\epsilon^2 + \chi^{(3)}\epsilon^3 + \cdots\right) \ldots\ldots\ldots\ldots\ldots\ldots\ldots\ldots(1)$$

Where $\epsilon$ is the applied external electric field.

In natural materials, higher order nonlinear susceptibility $\chi^{(2)}$, $\chi^{(3)}$, $\chi^{(4)}$ .. are extemely small[22] and hence nonlinear polarization is typically very weak. In this letter, we present a nonlinear metamaterial with enhanced second harmonic signal using a modified Split ring resonator (SRR) design. The modification of the split ring resonator is created by adding another ring as shown in Fig 1(c) which is a Double Split Ring Resonator (DSRR) with a shared varactor diode. Inserting a varactor diode to the design provides the nonlinearity to the unit cell.

Fig 1(a) shows the transmission through SRR and DSRR where SRR has a fundamental resonance frequency at $\boldsymbol{f_1}$ = 822MHz and DSRR has two distinct peaks with fundamental resonance frequencies at $\boldsymbol{f_1}$ = 825MHz and $\boldsymbol{f_2}$ = 1670MHz corresponding to two rings. The current density of the DSRR unit cell at the resonance frequencies are shown near each resonance frequency, which shows a LC resonance mode for the fundamental frequencies. Fig 1(b) shows the relationship between the inner ring frequency and outer ring frequency by plotting outer ring frequency with half of the inner ring frequency.

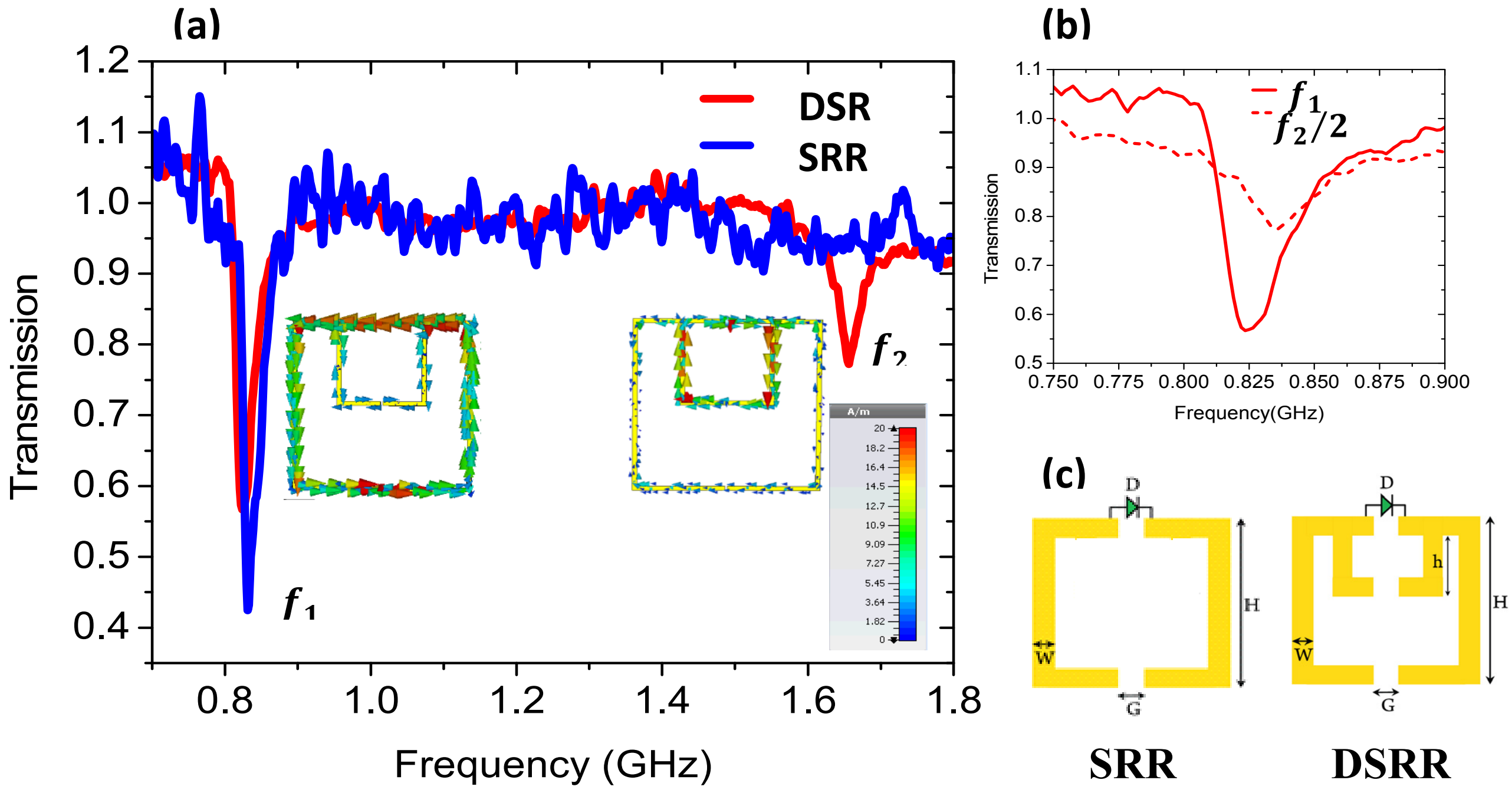

Figure 1 (a) Transmission for DSRR and SRR. (b) Alignment of inner ring and outer ring resonances for DSRR. (c) SRR and DSRR design

A varactor loaded split ring resonator can be moddeled as a RLC circuit as shown in fig 2(a) and fig 2(b) and in DSRR there is an extra capacitance added to the circuit in parallel to the exsiting capacitance due to the second ring in the structure.

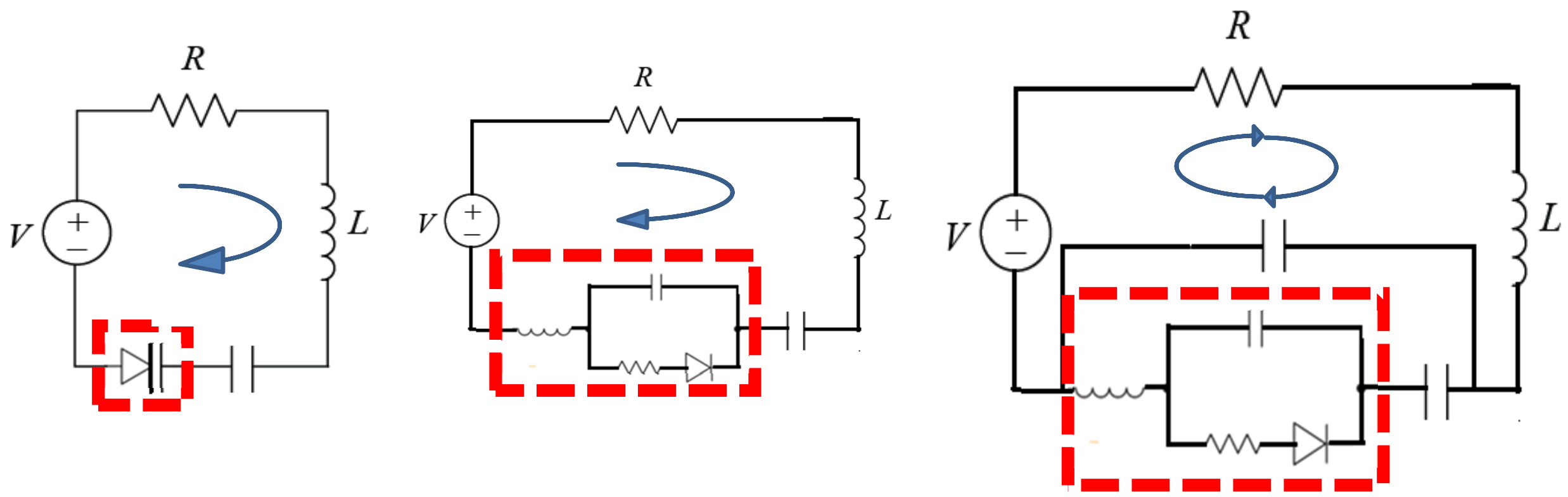

Figure 2 (a) Equivalent circuit of nonlinear SRR with diode. (b)&(c) Equivalent circuit of SRR (b) and DSRR (c) with effective circuit of diode.

In varactor loaded DSRR, the external signal provides the driven source which can be modeled as a virtual voltage source in the equavalent cicuit model.

$$V_{external}(t) = V_0 \cos \omega t$$

$$L\frac{dI}{dt} + RI + V_C + V_D = V_0 \cos \omega t$$

Where L is the total inductance of the DSRR, including self-inductance and mutual inductance, R is the effective resistance of the unit cell and $V_D$ is the voltage across the varactor diode.

Varactor loaded split ring resonator structure produces a second harmonic signal corresponding to its fundamental frequency $\omega_0 = 1/\sqrt{LC}$ . Since the capacitance of the varactor diode depends on the reverse bias voltage across the diode is given by the following expression,

$$V_D = \frac{1}{C}\int_0^t I\,dt = V_P\left[1 - \left(1 - \frac{Q}{C_\circ}\frac{1-M}{V_P}\right)^{1/(1-M)}\right]$$
$$= q + aq^2 + bq^3 + cq^4 + \cdots$$

Where higher order terms contributes to the harmonics generated by the structure. In the letter we only consider second harmonic generated by the varactor loaded SRR and DSRR structures. The amplitude of the second harmonic signal can be obtained using the circuit model of the SRR with the use of perturbation method under weak nonlinearity[23] and by considering only fundamental and second order terms.

The amplitude of the oscillations are given by

$$\tilde{q}(2\omega) = \frac{\alpha {V_0}^2}{\omega^3 Z(2\omega) Z(\omega)^2} \quad \text{.................................... (2)}$$

When the circuit resonates at $\omega = \omega_0$ and the impedance $Z(\omega)$ will be minimized resulting an increase of amplitude of the oscillations. In similar manner by minimizing $Z(2\omega)$ will further enhance the amplitude of oscillations and enhance the effect of second harmonic generation in the metamaterial.

In order to maximize the second harmonic and hence maximize other higher order harmonics, DSRR structure was fabricated such that outer ring size is 40mm x 40mm and inner ring size is 20mm x 20mm. Both gaps in two rings have a gap size of 1mm and the ring width is 1mm. When the inner ring size changes, it effects the second harmonic signal as shown in fig 3b and we observe maximum second harmonic power when the outer ring frequency ($f_1$) is twice as the inner ring frequency ($f_2$) as illustrated in Fig. 3a. However when the condition $f_2 = 2\, f_1$ does not satisfy, for 20mm and 22mm inner ring sizes we see two peaks in second harmonic signal and the power of the signal gets weaker as shown in Fig. 2b. In the case of satisfied condition of $f_2 = 2\, f_1$ , we have an enhancement of 70 when we compare a single split ring with a double split ring as. Furthermore, it can be seen that the probe signal power also can be used to increase second harmonic signal power and throughout this experiment we have used 0dBm(1mW) as the probe signal power.

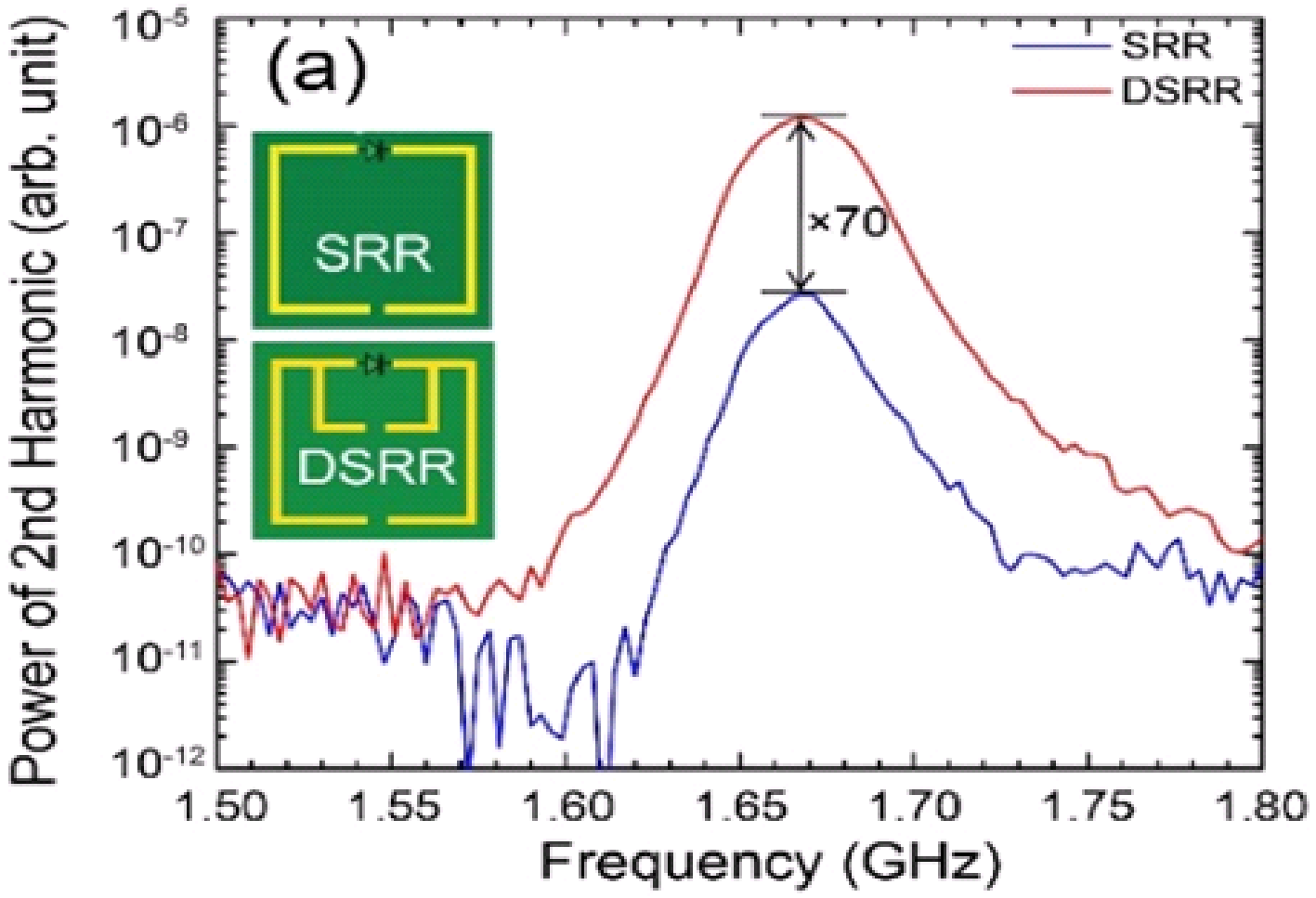

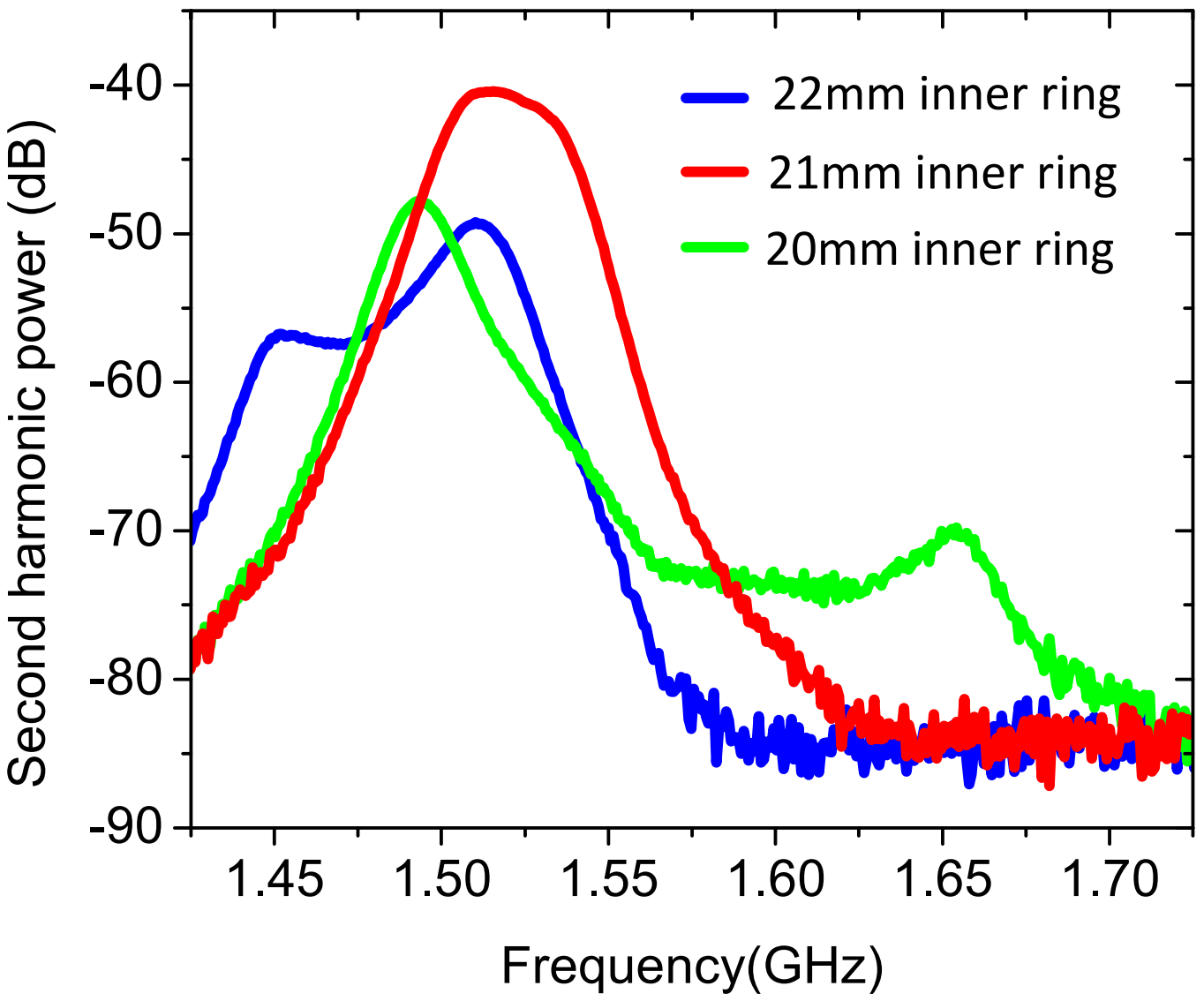

22mm inner ring
21mm inner ring
20mm inner ring
Second harmonic power (dB)
Frequency(GHz)

Figure 3 (a) SHG by SRR and DSRR. (b) SHG by DSRR with different inner ring size.

These varactor loaded nonlinear metamaterial unit cells can be considered as point sources of second harmonic signals which we demonstrate that these point sources has constructive and destructive interference when we keep them at specific positions.

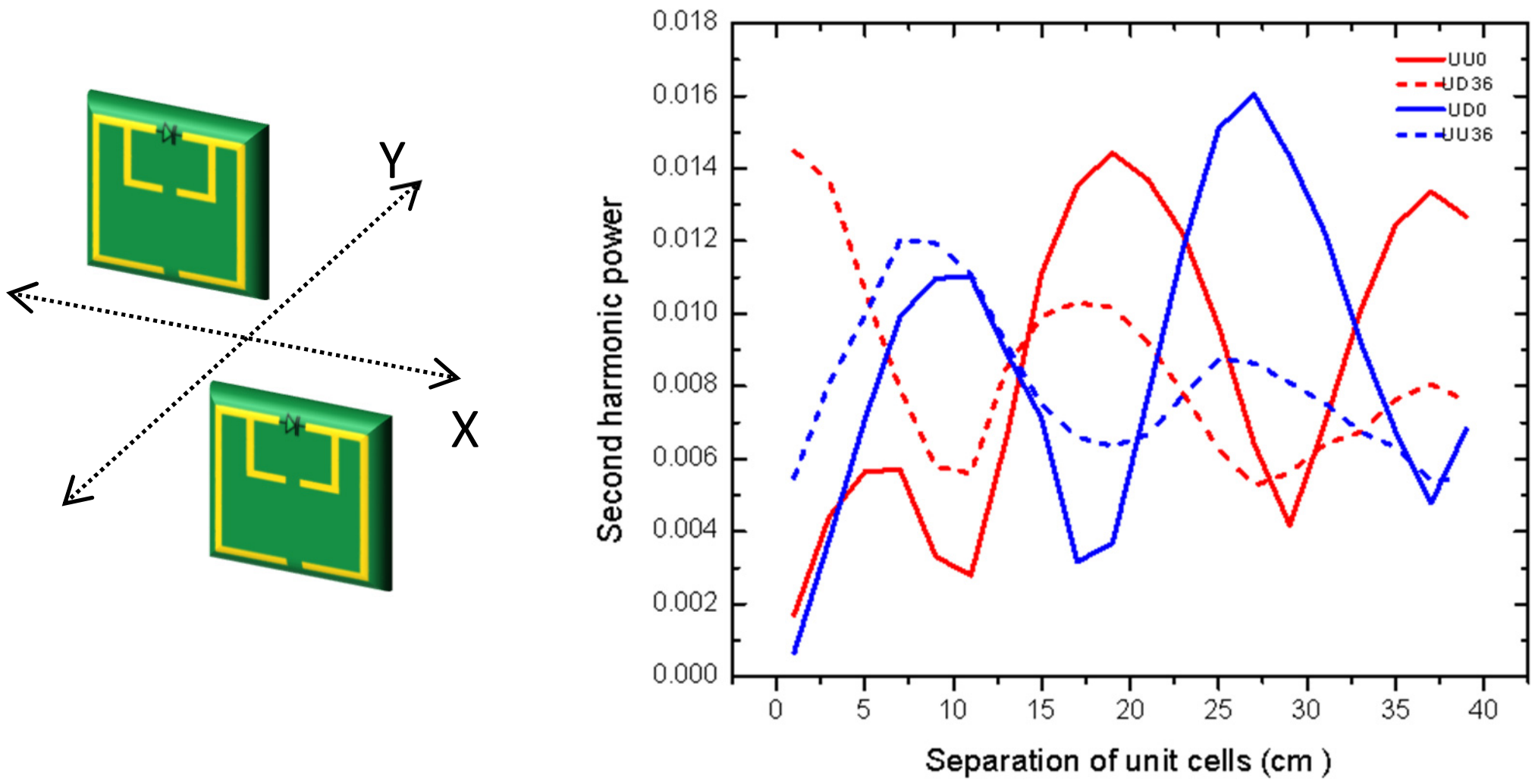

Figure 4 Measured SHG by two DSRRs.

Two unit cells of similar DSRR structures were arranged as shown in fig 3a and excited by a horn antenna from the direction indicated by x-axis and the second harmonic signal was collected from the direction of y using another horn antenna. Fig 3b shows the interference pattern when we change the relative distance in the y direction of the two unit cells. Second harmonic signal emits from the DSRR structure is 1668MHz, which corresponds to a wavelength of 17.98cm. In Fig3b solid line indicates the situation where the separation of the unit cells in x direction is 0cm and dashed line indicates that the separation of the unit cells in x direction is 36cm. in both cases relative change is in the y direction, and the dashed line represents the x distance between the unit cells is 36cm. The figure shows two different orientations with respect to varactor diode. One with the same orientation of the varactor diode in both unit cells and the other is 180 differences between the varactor diodes. Equation 3 gives the path difference of the two coherent second harmonic signals generated by 2 DSRR unit cells

$$\Delta l = \frac{(2l_y - l_x)}{\lambda_F} 2\pi \ldots\ldots\ldots\ldots\ldots\ldots\ldots\ldots\ldots\ldots\ldots\ldots\ldots (3)$$

Where $\Delta l$the path difference of the two signals is generated by DSRR cells, $l_y$ is the distance between the unit cells in y direction, $l_x$ is the distance between the unit cells in x direction and $\lambda_F$ is the wavelength of the fundamental frequency.

The fundamental wavelength of the signal is around 36cm and the second harmonic wavelength is 18cm. when separation of unit cells in x direction is either 0cm or 36cm, we can obtain the same interference pattern for both orientations of the diode as shown in Fig 3b (red solid line and dashed line indicates both cases). When the diodes are in the same orientation for λ/2 (9cm) we observe destructive interference and for λ (18cm) we observe a constructive interference. Similarly, for the opposite orientation of the diodes we observe constructive interference at a distance of λ/2 (9cm) and destructive interference at λ (18cm) which agrees with the Eq 3.

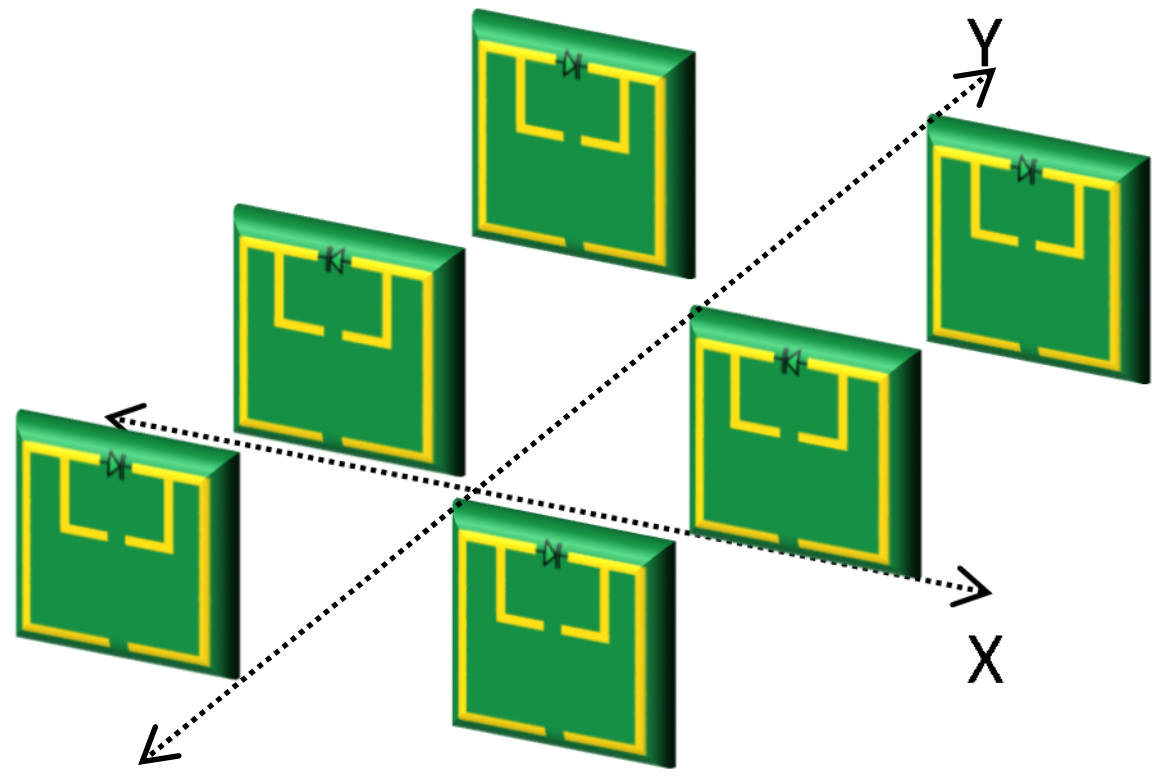

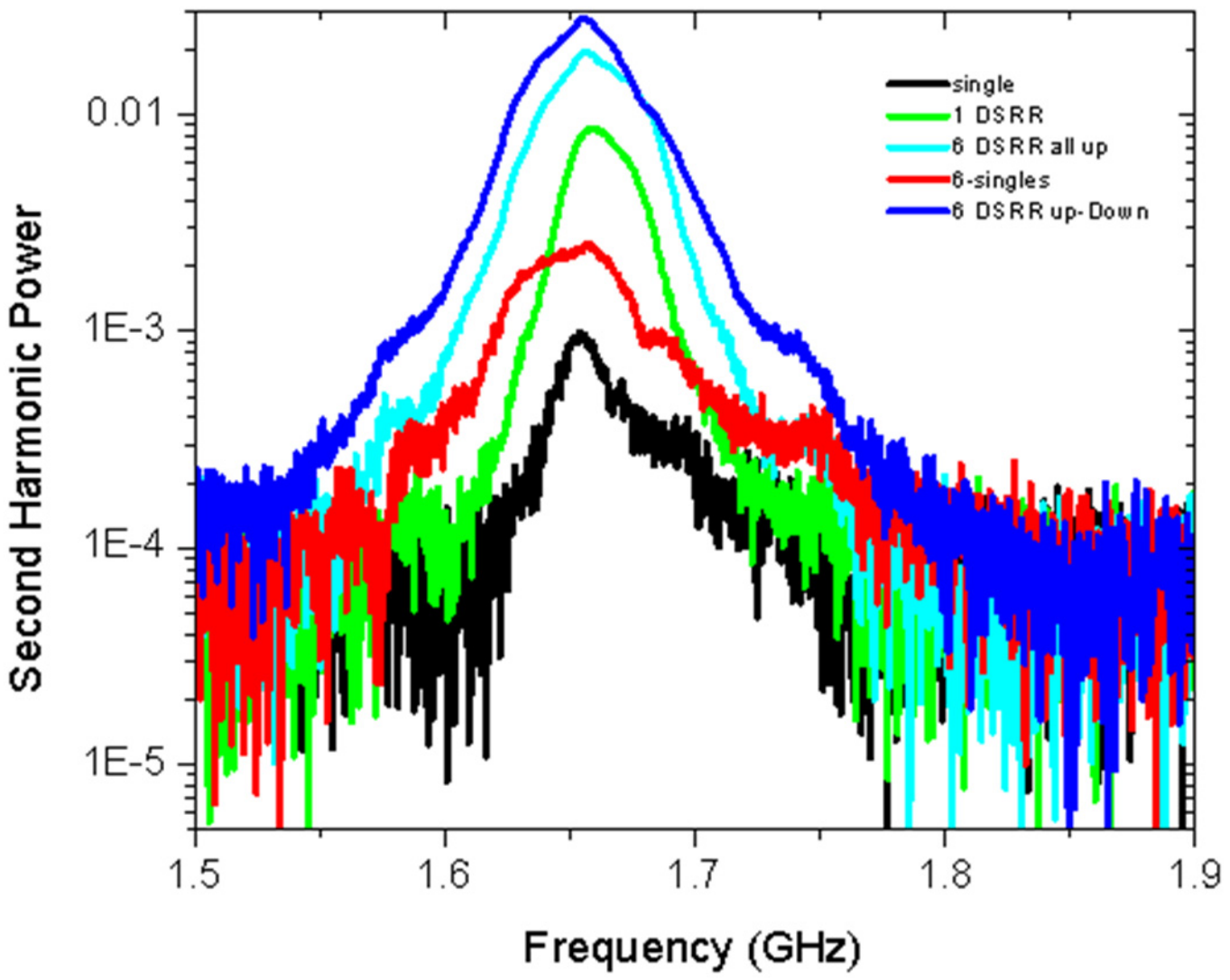

Figure 5 Array of DSRRs.

Studies have shown that for non-symmetric structures, second harmonic signal originates from the side arms and they are in phase. The SRR and DSRR structures are also non-symmetric and its second harmonic field is produced in the side arms of the structures are radiates to the far field.

By arranging unit cells in an array as shown in Fig 5a, can increase the total second harmonic signal with constructive interference of second harmonic waves from individual unit cells. Fig 5b shows interference of second harmonic signals from 6 unit cells of SRR and DSRR compared with their respective single unit cells. Distance between the unit cells in y direction $l_y$= 9cm and distance between the unit cells in x direction $l_x$ = 36cm and the direction of the diodes of the two unit cells in the middle is kept opposite to the other unit cells. Array is constructed such that in x direction unit cell orientation is in the same direction and in y direction unit cell direction is in opposite direction. When we compare the second harmonic signal power generated by six DSRR unit cells to the power of 1 DSRR unit cell the enhancement factor is 8. It indicates that not only interference of the second harmonic generated by DSRR unit cells is effecting the enhancement in the array structure but also these unit cells act as mirrors when they are at resonance and form a cavity inside the array.